\documentstyle[aps,multicol,epsf,rotate,citesort]{revtex}
\newcommand{\gl}[1]{Eq. (\ref{#1})}
\newcommand{\gls}[2]{Eqs. (\ref{#1},\ref{#2})}
\newcommand{\glto}[2]{Eqs. (\ref{#1}) to (\ref{#2})}
\def\gtrless{\raise2.5pt\hbox{$>$}\llap{\lower2.5pt\hbox{$<$}}}
\def\gtrapprox{\raise2.5pt\hbox{$>$}\llap{\lower2.5pt\hbox{$\approx$}}}
\newcommand{\bsq}[1]{\begin{subequations}\label{#1}}
\newcommand{\esq}{\end{subequations}}
\newcommand{\beq}[1]{\begin{equation}\label{#1}}
\newcommand{\eeq}{\end{equation}}
\newcommand{\beqa}[1]{\begin{eqnarray}\label{#1}}
\newcommand{\eeqa}{\end{eqnarray}} \newcommand{\fur}{\qquad\mbox{for
}\, } 
\newcommand{\kubo}[2]{\left( #1 | #2 \right)}
\newcommand{\kuboz}[3]{\left( #1 | #2 | #3 \right)}
\newcommand{\vek}[1]{{\bf #1}} \newcommand{\ii}{{ i}}

\begin{document}

\title{Light scattering spectra of supercooled molecular liquids}
\author{T.~Franosch$^{(a,b)}$, M.~Fuchs$^{(a)}$ and A.~Latz$^{(c)}$ }
\address{$^{(a)}$ Physik-Department, Technische Universit{\"a}t
M{\"u}nchen, D-85747 Garching, Germany;\\ $^{(b)}$ present address:
Lyman Laboratory of Physics, Harvard University, Cambridge,
Massachusetts 02138\\ $^{(c)}$ Institut f{\"u}r Physik, Johannes
Gutenberg Universit{\"a}t, Staudinger Weg 7, D-55099 Mainz}
\date{\today} \maketitle

\begin{abstract}
The light scattering spectra of molecular liquids are derived within a
generalized hydrodynamics. The wave vector and scattering angle
dependences are given in the most general case and the change of the
spectral features from liquid to solidlike is discussed without
phenomenological model assumptions for (general) dielectric systems
without long--ranged order.  Exact microscopic expressions are derived for
the frequency-dependent transport kernels, generalized thermodynamic
derivatives and the background spectra.  \bigskip

 \medskip

\end{abstract}

\begin{multicols}{2}

\section{Introduction}
 
Light scattering is a powerful tool to study the dynamics of dense
(transparent) materials \cite{Berne76}. The fluctuations of the
dielectric tensor for wave vector transfer $\vek{q}$ are measured,
where the corresponding wavelength can be considered large compared to
molecular length scales.  In this case the theoretical description of
the light spectra simplifies as it suffices to determine the lowest
orders in wave vector $\vek{q}$ only.  The wavevector and
scattering--angle dependence of the scattering cross sections thus can
be determined.

Focusing on low frequencies, the hydrodynamic approach can be used to
calculate the spectral shapes.  For the polarized light scattering
spectra of gases and liquids, the assumption that dielectric
fluctuations are dominantly caused by density fluctuations leads to
the well known Rayleigh--Brillouin spectra. The corresponding
hydrodynamic depolarized spectra were first obtained within simplified
models by Andersen and Pecora \cite{Andersen71} and Keyes and Kivelson
\cite{Keyes71}. In the latter case, upon the assumption that
fluctuations in the orientations of the molecules cause the dielectric
variations. As observed experimentally in molecular liquids, a
negative central line appears called ``Rytov--dip''
\cite{Starunov67,Stegeman68}.

The spectra, in particular the depolarized ones, change qualitatively
if the relaxation times of the structural dynamics increase upon
cooling the liquids. For the polarized spectra Mountain introduced a
frequency dependent longitudinal viscosity in order to model the
additionally appearing central line \cite{Mountain66}. In the
depolarized light scattering spectra transverse sound peaks become
visible as expected from hydrodynamic calculations for solids
\cite{Landau8}.  Within phenomenological models including
non--hydrodynamic variables, the changes of the spectra from
liquidlike at high to solidlike at low temperatures could be explained
\cite{Vaucamps,Quentrec,Wang1,Chapell,Drey1,Drey2}.  However,
assumptions about the included slow variables and about
phenomenological kinetic equations coupling their time dependences
were required. The introduction of memory functions, in most detail in
the recent work of Dreyfus et al.  \cite{Drey1,Drey2}, has relaxed the
requirement to identify and include all slow variables but still uses
phenomenological equations to couple rotational and translational
degrees of freedom.

Here, we clarify that generalized hydrodynamics and symmetry
considerations suffice to 
explain the light scattering spectra
qualitatively as they change from liquid-- to solidlike. The results
will be derived without any assumptions about non--hydrodynamic
variables (and their couplings) and will also not depend on specific
light scattering mechanisms nor molecular parameters like shape,
dipole--moment, polarizabilities nor chirality.  The set of slow
variables we consider are the standard slow hydrodynamic variables of
liquids, density $n(\vek{q})$, current density $\vek{j}(\vek{q})$, and
temperature $\Theta(\vek{q})$ (connected to energy conservation), and
the slow structural relaxation of supercooled liquids enters via a few
memory functions.  We thus provide a general framework for the
analysis of light scattering spectra in supercooled molecular liquids
which we expect will prove useful either for phenomenological
discussions using fit--functions for the memory kernels --- we will
list all restrictions on these fit--functions --- or for the
consideration of specific scattering mechanisms.  Our central
technique for simplifying the spectra consists in small wave vector
expansions of the memory functions as should be appropriate for
disordered systems. Thus we adopt the idea of generalized
hydrodynamics, which extends the regular hydrodynamic approach to
larger frequencies. In detail, we use the one suggested by G\"otze and
Latz \cite{Goetze89b} as it provides a physically reasonable
description of glassy systems.  Finally, we also derive Green--Kubo
formulae  which enable e.~g. computer
simulations to determine the memory functions and thus the complete
spectra directly.

Our assumptions pertain to the systems under consideration and can be
tested experimentally: We consider only the lowest non--trivial orders
in wave vector transfer $q$, in order to find general results for the
light scattering from amorphous dielectric and macroscopically
isotropic and optically inactive materials within the framework of
linear response, classical statistical mechanics and classical
electromagnetism. The condition $qa\ll1$, where $a$ denotes either a
typical molecular size, the average particle distance or a collective
correlation length, appears well satisfied for supercooled molecular
liquids but excludes studies of critical phenomena. Electromagnetic
retardation also can be neglected for $\omega\ll c q$, where $c$ is
the speed of light.

The general formulae for the spectra and constitutive equations are
presented in Section \ref{section2}. Section \ref{section3} lists the
central results which are discussed in Section \ref{section4}. The
results for the depolarized spectra are compared to previous
theoretical approaches in Section \ref{section5}, and conclusions in
Section \ref{section6} summarize our results. More technical aspects
are contained in Appendizes A, B and C, and Appendix D outlines the
application of our general results to specific light scattering
mechanisms.

\section{General formulae}
\label{section2}

\subsection{Dielectric fluctuations}

The dielectric permeability fluctuates in space and
time around its average and additionally a (diffusive) scattering
occurs. In general, this scattering spectrum reflects the dynamical
processes in the sample and depends on the frequency of the incident,
$\omega_i$, as well as the scattered wave, $\omega_f$. A simplifcation
is possible if one considers only small frequency shifts,
$\omega=\omega_i-\omega_f$ with $|\omega|\ll\omega_i$.  Then dynamical
dielectric fluctuations
$\kubo{\epsilon_{ij}(\vek{q},t)}{\epsilon_{kl}(\vek{q})}$ determine
the scattering cross sections completely \cite{Landau8}. Here the Kubo
scalar product, $\kubo{A(t)}{B}=(1/k_BT) \langle \delta A^*(t) \delta
B\rangle$, is used, with $T$ temperature and $k_B$ Boltzmann's
constant.

The fluctuation $\delta \epsilon_{ij}(\vek{q},t)$ has even time
reversal symmetry, is a symmetric tensor of second rank, and, as
Fourier-transform of a real quantity, fulfills
$\epsilon_{ij}(-\vek{q},t) = \epsilon_{ij}(\vek{q},t)^*$. In
particular, the long-wave-length limit $\epsilon_{ij}(\vek{q} \to
0,t)$ is real.  Different Cartesian components $i,j,k,l$ are picked
out depending on the polarization directions of the incoming and
scattered light \cite{Berne76}; see Appendix A for more details.  The
dynamical evolution is given by the Liouvillian $\cal L$ via
$\partial_t A = i{\cal L} A$.  A Laplace transformation --- convention
$f(z) = i \int_0^\infty dt e^{izt} f(t)$ for $\Im z>0$ --- thus leads
to the problem to calculate, for $q\to0$, the matrix elements of:
\begin{equation}
\kubo{\epsilon_{ij}(\vek{q},z)}{\epsilon_{kl}(\vek{q})} =
\label{e1}\left( \epsilon_{ij}(\vek{q}) | \frac{1}{{\cal L}-z} | 
\epsilon_{kl}(\vek{q}) \right)\; .
\end{equation}
The spectra at frequency $\omega$ then are given by the imaginary part
of $\kubo{\epsilon_{ij}(\vek{q},\omega+ i0)}{\epsilon_{kl}(\vek{q})}$
denoted by
$\kubo{\epsilon_{ij}(\vek{q},\omega)}{\epsilon_{kl}(\vek{q})}''$.
They are real, positive and symmetric functions of frequency.

\subsection{Generalized hydrodynamics}

Light scattering measures large wavelength dielectric fluctuations.
Even though $qa\ll1$ can thus be assumed, the limit $q\to0$ cannot be
performed naively in \gl{e1}. Because of density--, momentum--, and
energy--conservation, there are poles in the resolvent $R(z)=({\cal
L}-z)^{-1}$, which shift to vanishing frequency in this limit
\cite{Forster75}. Using the Zwanzig--Mori formalism, these
hydrodynamic low frequency features can be identified. One introduces
 the reduced
resolvent 
\begin{equation}\label{e2}
R'(z) = Q \frac{1}{Q{\cal L}Q-z} Q \; ,
\end{equation}
where the projector $Q=1-P$ projects perpendicular to the hydrodynamic
modes:
\begin{equation}\label{e3} 
P = \frac{ | n(\vek{q}) ) (n(\vek{q}) |}{ (n(q) | n(q))} + \frac{|
\Theta(\vek{q}) ) (\Theta(\vek{q}) | }{(\Theta(q) | \Theta(q) )} +
\sum_i \frac{ | j_i(\vek{q}) ) ( j_i(\vek{q}) | }{(j_i(q)|j_i(q))}
\end{equation}
The standard hydrodynamic formulae are obtained in this approach if
the reduced resolvent $R'(z)$ is treated in a Markovian approximation,
replacing its matrix elements with frequency independent transport
coefficients \cite{Forster75}. Generalized hydrodynamics differs from
this by retaining the frequency dependence of $R'(z)$ but still
neglecting its wave vector dependence. This generalization is required
for liquids at lower temperatures as the structural relaxation slows
down strongly.

Following Ref. \cite{Goetze89b} we identify the fluctuating
 temperature $\Theta(\vek{q})$ with the kinetic energy fluctuations,
 $e^K(\vek{q})$, that are orthogonal to the density fluctuations,
 $c_V^0 \Theta(\vek{q}) = Q_n e^K(q) = e^K(\vek{q}) - n(\vek{q}) (e^K
 | n)/(n | n)$. Here $c_V^0 = 3 k_B/2$ abbreviates the specific heat
 per particle of the kinetic degrees of freedom, and $Q_n$ is the
 projector orthogonal to the density.  Conservation of the total
 energy $e(\vek{q}) = e^K(\vek{q}) + e^P(\vek{q})$ implies
\begin{equation}
\label{energy}
c_V^0 {\cal L} \Theta(\vek{q}) = q j^{L}_e(\vek{q}) - q j^L(\vek{q})
\frac{(e|n)}{(n|n)} - {\cal L} Q e^P(\vek{q}) \, ,
\end{equation}
 where superscripts $L$ indicate the longitudinal part, $j^L =
\vek{q}\cdot \vek{j} / q$, $e^P(\vek{q})$ is the potential energy, and
$\vek{j}_e(\vek{q})$ the total energy current. Note that since
$(\Theta(\vek{q}) | e^P(\vek{q})) = 0$ one can replace $Q_n
e^P(\vek{q}) = Q e^P(\vek{q})$.  The hydrodynamic variables are
orthogonal with normalizations:
$\kubo{n(\vek{q})}{n(\vek{q})}=NS(q)/k_BT$,
$\kubo{j_k(\vek{q})}{j_l(\vek{q})} = (N/m) \delta_{kl}$ and
$\kubo{\Theta(\vek{q})}{\Theta(\vek{q})} = NT/c_V^0$.  Here, $n$ is
the average density of $N$ molecules, $m$ the molecular mass, and
$S(q)$ is the equilibrium center--of--mass structure factor;
$\delta_{kl}$ is Kronecker's symbol.
Throughout the following we will neglect wave vector dependences
caused by molecular length scales, and replace, e. g., the structure
factor by its homogeneous limit given by the isothermal
compressibility $\kappa_T$: $S(q)=S(0)+{\cal O}((qa)^2)$ with
$S(0)=nk_BT \kappa_T$.

Considering, in generalized hydrodynamics \cite{hansen}, the
 fluctuating temperature instead of energy fluctuations rests upon the
 experimental observation that the heat conduction of glasses and
 liquids is not drastically different.  This aspect is discussed in
 Refs.  \cite{Goetze89b,Latz99} where the generalized hydrodynamics
 also is tested by molecular dynamics simulations. This formulation of
 generalized hydrodynamics accounts straightforwardly for a frequency
dependent isochoric specific heat.

\subsection{Decomposition of dielectric fluctuations}

The exact resolvent calculus sketched in the previous section thus
  provides a reformulation of \gl{e1}, see Eqn. (A14) in
  Ref. \cite{Goetze89b}.  The reduced dynamics $R'(z)$ and the
  projector $P$, which projects onto the hydrodynamic variables,
  appear:
\begin{eqnarray}\label{e4}
(\epsilon_{ij}(\vek{q}) | R(z) | \epsilon_{kl}(\vek{q}) ) =
(\epsilon_{ij}(\vek{q})| R'(z) | \epsilon_{kl}(\vek{q}) ) + \nonumber
\\ (\epsilon_{ij}(\vek{q}) | (1 - R'(z) {\cal L} ) P R(z) P (1-{\cal
L} R'(z) ) | \epsilon_{kl}(\vek{q}) )
\end{eqnarray}
Thus the coupling of the dielectric fluctuations to the hydrodynamic
variables is found; explicitly it is given when writing out $PR(z)P$
in \gl{e4}.  Additionally there is a background spectrum, the first
term on the right hand side of \gl{e4}.

Since the hydrodynamic modes have been projected out, (generalized)
hydrodynamics postulates that the limit $q \to 0$ in $R'(z)$ can now
be performed safely.  
This
leads to the well--known result for Raman scattering \cite{Landau8}:
The background spectrum consists of scalar, and symmetric traceless
tensor scattering
\begin{eqnarray}\label{e5}
(\epsilon_{ij}(\vek{q}) | R'(z) | \epsilon_{kl}(\vek{q}) ) = {\cal
S}(z) \delta_{ij} \delta_{kl} + \nonumber \\  {\cal T}(z)
(\delta_{ik} \delta_{jl} + \delta_{il} \delta_{jk} -\frac{2}{3}
\delta_{ij} \delta_{kl} ) +{\cal O}(q^2) \; ,
\end{eqnarray} 
Explicit expression for ${\cal S}(z)$ and ${\cal T}(z)$ can be
obtained by choosing special linear combinations of the dielectric
tensor.  Let $s_{00} = [\epsilon_{xx} +
\epsilon_{yy}+\epsilon_{zz}]/3$ denote the long-wave-length limit of
the scalar part and $t_{20} = [2 \epsilon_{zz} -
\epsilon_{xx}-\epsilon_{yy}]/\sqrt{12}$ the $\nu=0$ component of the
corresponding spherical tensor $t_{2\nu}$; (the prefactors are
conventional).  Then, ${\cal S}(z) = (s_{00}| R'(z)|s_{00} )$ and
${\cal T}(z) = (t_{20}|R'(z)|t_{20})$ are the scalar and the tensorial
correlations;  
see Appendix B for a discussion of
the applicability of these considerations to polar fluids.
  Both spectral contributions ${\cal
S}(z), {\cal T}(z)$ are autocorrelation functions of real variables
with even time inversion symmetry. Their spectra thus are even and
non--negative.

In order to simplify the discussion of the couplings to the
hydrodynamic variables in \gl{e4}, it is useful to consider the
corresponding generalized constitutive equations
\cite{Forster75,Goetze89b}. These describe the temporal decay of the
deviations in a variable, say in $\delta \epsilon_{ij}(\vek{q},t)$,
produced by an adiabatic perturbation with external fields coupling to
the conserved variables, after the perturbations are switched of at
time $t=0$:
\begin{eqnarray}\label{e4b}
 \langle \delta \epsilon_{ij}(\vek{q},t) \rangle = \sum_\alpha^5\;\;
\langle \delta A_\alpha(t) \rangle\; (A_\alpha |
\epsilon_{ij}(\vek{q}))/ (A_\alpha | A_\alpha ) -\nonumber\\ \int_0^t
d\tau \langle \delta A_\alpha(\tau) \rangle\; ( A_\alpha | i {\cal L}
R'(t-\tau) | \epsilon_{ij}(\vek{q})) / (A_\alpha | A_\alpha ) \; ,
\end{eqnarray}
where the (orthogonal) hydrodynamic variables are abbreviated: $A_1 =
n(\vek{q}), A_2 = \Theta(\vek{q}), A_3 = j_x(\vek{q}), A_4 =
j_y(\vek{q}), A_5 = j_z(\vek{q}))$.  The special choice of the
external perturbation considered in the constitutive equation,
\gl{e4b}, prepares a fluctuation in $ \langle \delta
\epsilon_{ij}(\vek{q},t) \rangle $ which decays slowly because it
requires decay of fluctuations of the conserved variables at large
wavelengths, ${\cal O}(1/q)$. Thus the time evolution of $ \langle
\delta \epsilon_{ij}(\vek{q},t) \rangle $ in \gl{e4b}, is determined
by time-dependent couplings to the generalized hydrodynamics of the
distinguished variables.  As seen trivially via a
Laplace--transformation, the identical couplings appear in \gl{e4} as
in \gl{e4b}, expressing that via these couplings, close to
equilibrium, dielectric fluctuations at long wavelengths acquire slow
hydrodynamic components. Equation (\ref{e4}) further identifies the
non--hydrodynamic components contributing to the background.  The
first term on the right hand side of \gl{e4b} describes static or
instantaneous couplings, whereas the second term describes dynamic
couplings which need time to build up and may be characterized by a
finite response time. These will become important when approaching the
glass transition upon cooling as then the structural relaxation times
increase.

Density fluctuations are coupled to the dielectric fluctuations
statically via the scalar scattering mechanism only, as follows from
\gl{e4b} when inserting the projector $P$ from \gl{e3}:
\begin{equation}\label{e6}
(n(\vek{q}) | (1 - {\cal L} R'(z) ) | \epsilon_{ij}(\vek{q}) ) = (n |
s_{00} ) \delta_{ij} +{\cal O}(q^2)\; .
\end{equation}
The dynamic coupling vanishes because ${\cal L} n(\vek{q}) = q_k
j_k(\vek{q})$ is again a hydrodynamic variable.  Furthermore the
scalar density cannot couple to the dielectric tensor fluctuations
$t_{2\nu}$ in the limit $q\to0$.

Similar arguments hold for the coupling of the temperature to the
 dielectric fluctuations. Since the kinetic energy is not conserved,
 there is a dynamic coupling in addition to the static one. In order
 to guarantee conservation of the total energy \gl{energy} is
 used. Observing $ ({\cal L} Q e^P | R'(z) = (Q e^P| + z (e^P | R'(z)$
 and rearranging terms one finds to lowest order in $q$.
\begin{eqnarray}\label{e7}
c_V^0(\Theta(\vek{q}) | (1 - {\cal L} R'(z) ) | \epsilon_{ij}(\vek{q})
) = \nonumber\\ \delta_{ij} [(Q_n e |s_{00}) + z ( e^P | R'(z)
|s_{00})] + {\cal O}(q)\; .
 \end{eqnarray}
  Further couplings of order $q$ can be ignored.  The dynamic coupling
in \gl{e7} arises from the separation of the total energy fluctuations
into fast kinetic and (possibly) slow potential ones. Thus the
simplified handling of the reduced resolvent in this generalized
hydrodynamics has to be paid by an additional frequency dependent
coupling to temperature fluctuations.

Finally for the coupling of dielectric fluctuations to current
fluctuations
\begin{eqnarray}\label{e8}
& (j_k(\vek{q}) | (1\! -\! {\cal L} R'(z) ) | \epsilon_{ij} (\vek{q})
) = & \nonumber\\ & - \sum_l q_l (\tau_{kl}(\vek{q}) | R'(z) |
\epsilon_{ij}(\vek{q}) )/m + {\cal O}(q^2)\; , &
\end{eqnarray}
one finds a purely dynamical coupling, as first recognized within
so--called two variable models \cite{Andersen71,Keyes71}.  A static
coupling is excluded, since the currents $j_k$ and the dielectric
tensor $\epsilon_{ij}$ have different time reversal symmetry.  The
dynamic coupling to the stress tensor, $\tau_{kl}$, which appears
because of momentum conservation, ${\cal L} {j}_k(\vek{q})= \sum_l q_l
\tau_{kl}(\vek{q})/m$, can be evaluated in the long-wavelength limit:
\begin{eqnarray}\label{e9}
(\tau_{kl}(\vek{q}) | R'(z) | \epsilon_{ij}(\vek{q}) ) = (p \; |
R'(z)| s_{00} ) \delta_{ij} \delta_{kl} + \nonumber \\ 
(\tau_{20}|R'(z)|t_{20}) (\delta_{ik} \delta_{jl} + \delta_{il}
\delta_{jk} -\frac{2}{3} \delta_{ij} \delta_{kl} ) + {\cal O}(q^2)\; .
\end{eqnarray}
Here $p= [\tau_{xx}+\tau_{yy}+\tau_{zz}]/3$ and $\tau_{20} = [2
\tau_{zz}-\tau_{xx}-\tau_{yy}]/\sqrt{12}$ denote the long-wavelength
limits of the pressure and
of the transversal stress tensor.

According to the basic assumption of (generalized) hydrodynamics
\cite{Forster75,Goetze89b}, one has to keep terms only to the order
indicated, the remaining ones are assumed to be regular with respect
to frequency $z$ in the limit $q \to 0$.

\section{Results}
\label{section3}

From Eqs. (\ref{e4}) to (\ref{e9}) the light scattering spectra follow
if a scattering geometry is chosen and the appropriate tensor elements
are calculated; see Appendix A for the used geometry.  Polarizations
vertical to, $V$, or in the scattering plane, $H$, are considered,
where the standard abbreviation $I_{io}$ denotes polarizations for
incoming and outgoing light.  The spectra depend on $q$, $\omega$ and
scattering angle $\theta$.  We find
$I_{HV}(q,\theta,\omega)=I_{VH}(q,\theta, \omega)$ as predicted by
Rayleigh's reciprocity theorem \cite{Berne76}.

\subsection{Total scattered intensities}

The total scattered intensities (except for standard prefactors
\cite{Berne76}) can be obtained directly from \gl{total} in Appendix A
and consist of scalar and tensorial scattering \cite{Landau8}:
\begin{equation}\label{e9b}
  I_{VV}(q,\theta) = \left( s_{00} | s_{00} \right) +\frac{1}{3}
  \left(t_{20} | t_{20} \right)\; ,
\end{equation}
\begin{equation}\label{e9c}
  I_{VH}(q,\theta) =  \left(t_{20} | t_{20} \right) \; ,
\end{equation}
\begin{eqnarray}\label{e9d}
  I_{HH}(q,\theta) = \cos^2 \theta \left( s_{00} | s_{00} \right) +
 \nonumber\\ ( 1 + \frac{1}{3} \cos^2 \theta) \left(t_{20} |
 t_{20} \right) \; .
\end{eqnarray}
The intensities are wave vector independent as the limit $qa\ll1$ is
considered in systems where all molecular correlations are
short--ranged.  The conservation laws affect the spectral shapes only,
i.e. cause low lying hydrodynamic lines, but do not lead to
long--ranged static correlations.  Concurrently, the static couplings
of the conserved variables to the dielectric fluctuations in
\glto{e6}{e8} are $q$--independent for small wave numbers.  See
Section IV, why these results appear violated when considering the
hydrodynamic limits.

\subsection{Depolarized spectrum}

Using the decomposition of the offdiagonal dielectric fluctuations,
Eqs. (\ref{e4b}), (\ref{e8}) and (\ref{e9}), one can identify a
dynamic coefficient, $a_{VH}(z)$, which describes the coupling to the
generalized hydrodynamic variables:
\begin{equation}\label{e10}
  a_{VH}(t) = \frac{1}{N} \left( \tau_{20} | R'(t) |t_{20} \right) \;
  .
\end{equation}
It is a generalized elasto--optic or Pockels' constant familiar from
light scattering in solids \cite{Landau8}, and is a real and symmetric
function of time, as the two tensor variables determining it are real
with even time parity. Its Laplace transform, $a_{VH}(z)$, therefore
has an even spectrum, $a_{VH}''(\omega)$. This Pockels' constant
describes the dynamic coupling of the transverse current into the
dielectric fluctuations and the constitutive equation becomes:
\beq{e10b} \langle \delta \epsilon_{VH}(\vek{q},t) \rangle = - i q
\cos \frac \theta2 \int_0^t d\tau\; a_{VH}(t-\tau) \langle \delta
j_y(\vek{q},\tau) \rangle \; . \eeq From \gls{e4}{e5} thus follows the
general result for the depolarized spectrum:
\begin{eqnarray}\label{e12} 
  \left( \epsilon_{VH}(\vek{q},z) | \epsilon_{VH}(\vek{q}) \right) =
   {\cal T}(z) + \nonumber\\ q^2 \cos^2 \frac{\theta}{2} \,
  a_{VH}(z)^2 C_{jj}^T(q,z) \; .
\end{eqnarray}
Here $C_{jj}^T(q,z) = \left( j^T(q,z) | j^T(q) \right)$ denotes the
correlation function of the transversal current fluctuations.  The
spectrum consists of a background arising from the symmetric
scattering in \gl{e5}, which commonly is discussed as Raman line, and
of couplings to the current fluctuations.  Naively evaluating the
depolarized spectrum at vanishing wave vector would neglect this
additional contribution. It is small, of order ${\cal O}(q^2)$, but is
characterized by a time scale that diverges in the hydrodynamic limit,
$q\to0$, and therefore dominates the low frequency spectrum.  The full
current correlators appear in \gl{e12}, which stresses that no
assumptions about translational--rotational coupling are required in
order to derive \gl{e12}. Explicitly this has been shown by the
derivation of \gl{e12} for a liquid of spherical particles in
Ref. \cite{Fuchs91}, which was tested in a simulation \cite{Madden}.

\subsection{Polarized Spectra}

For the $VV$ spectrum, \gl{e9} suggests to introduce the
elasto-optical constant
\begin{equation}\label{avv}
a_{VV}(t) = \frac 23\; a_{VH}(t) - \frac{ \left(p \, | R'(t) | s_{00}
\right)}{N} \; .
\end{equation}
It has identical properties as $a_{VH}(t)$ since the tensor variables
entering its definition again are real with even time parity.  Another
time dependent coupling function arises from the temperature
fluctuations as described in \gl{e7}:
\begin{equation}\label{xi}
\xi(z) = \left. \frac{\partial s_{00}}{\partial T} \! \right)_n + z
\frac{(e^P | R'(z)| s_{00} )}{N T} \; ,
\end{equation}
where the thermodynamic derivative $NT(\partial s_{00}/\partial T)_n=
\kubo{Q_ne}{s_{00}}$ is written, which contains the total energy
perpendicular to density fluctuations \cite{Goetze89b}.  Clearly,
\gl{xi} presents a generalized time or frequency dependent
thermodynamic derivative. The scalar variables entering its time or
frequency dependent term are real with even time parity, and
$\xi''(\omega)/\omega$ consequently is an even function of $\omega$.
The constitutive equation coupling the fluctuating hydrodynamic
variables into $\delta \epsilon_{VV}(\vek{q},t)$, from \gl{e4b},
becomes:
\begin{eqnarray}
\label{VVvariable}
& \langle \delta \epsilon_{VV}(\vek{q},z) \rangle = (\partial
s_{00}/\partial n)_T \langle \delta n(\vek{q},z) \rangle + & \nonumber
\\ & \xi(z) \langle \delta \Theta(\vek{q},z) \rangle + a_{VV}(z) q
\langle \delta j^L(\vek{q},z) \rangle \; ,&
\end{eqnarray}
where the thermodynamic relation $( n | s_{00}) = N n
\kappa_T(\partial s_{00}/\partial n)_T$ is used.  Collecting the terms
in \gl{e4} one obtains when applying the mass--conservation law, which
gives $q^2 C_{jj}^L(q,z)= q z C_{nj}^L(q,z) = z^2 C_{nn}(q,z)+ z N n
\kappa_T , q C_{\Theta j}^L(q,z) = z C_{\Theta n}(q,z)$,
\begin{eqnarray}
\label{VVspectrum}
& & (\epsilon_{VV}(\vek{q},z)| \epsilon_{VV}(\vek{q}) ) = {\cal S}(z)
 +4 {\cal T}(z)/3 \nonumber\\ & & + z a^2_{VV}(z) N n \kappa_T + 2
 (\partial s_{00}/\partial n)_T a_{VV}(z) N n \kappa_T \nonumber \\ &
 & + [ (\partial s_{00}/\partial n)_T + z a_{VV}(z)]^2 C_{nn}(q,z)
 \nonumber \\ & & + \xi(z)^2 C_{\Theta \Theta}(q,z) \nonumber \\ & &+
 2 [(\partial s_{00}/\partial n)_T +z a_{VV}(z)] \xi(z) C_{n
 \Theta}(q,z) \; .
\end{eqnarray}
Here $C_{nn}(q,z)$ denotes the density-density correlation function,
$C_{\Theta \Theta}(q,z)$ the temperature-temperature correlation
function, etc; see Appendix C.  Equation (\ref{VVspectrum}) is our
principal result for the VV spectrum.  It extends the conventional
hydrodynamic spectra to arbitrary frequencies. The small-wave-vector
singularities are encoded in the generalized hydrodynamic correlation
functions, $C_{\alpha\beta}(q,z)$, which are determined by the true
resolvent, $R(z)$, and can thus e.~g. be obtained from simulations.

From the information on the depolarized and the polarized spectrum
also the $I_{HH}$ spectrum can be obtained even though it is not a
simple linear combination.  The fluctuating variable coupling to the
distinguished variables for the $HH$ scattering in the geometry of
Appendix A is given by
\begin{eqnarray}
& & \langle \delta \epsilon_{HH}(\vek{q},z) \rangle = - \cos \theta \,
  (\partial s_{00}/\partial n )_T \langle \delta n(\vek{q},z) \rangle
  \nonumber \\ & & - \cos \theta \; \xi(z) \langle \delta
  \Theta(\vek{q},z) \rangle \nonumber \\ & & - [a_{HH}(z) \cos \theta
  - a_{VH}(z) ] q \langle \delta j^L(\vek{q},z) \rangle \; ,
\end{eqnarray}
where we abbreviated $a_{HH}(z) = a_{VV}(z) - a_{VH}(z)$.  The general
spectrum in the $HH$ geometry now reads
\begin{eqnarray}
\label{HHspectrum}
& &(\epsilon_{HH}(\vek{q},z) | \epsilon_{HH}(\vek{q})) =
{\cal S}(z) \cos^2 \theta + {\cal T}(z)(1 + \frac{1}{3} 
\cos^2 \theta) \nonumber \\ & & + z \left[ a_{HH}(z) \cos \theta -
a_{VH}(z) \right]^2 N n \kappa_T \nonumber \\ & & + 2
\left. \frac{\partial s_{00}}{\partial n} \! \right)_T \cos \theta
\left[ a_{HH}(z) \cos \theta - a_{VH}(z) \right] N n \kappa_T
\nonumber \\ & & + \left\{ \left[ \left. \frac{\partial
s_{00}}{\partial n} \! \right)_T + z a_{HH}(z) \right] \cos \theta - z
a_{VH}(z) \right\}^2 C_{nn}(q,z) \nonumber \\ & & + \xi(z)^2 \cos^2
\theta \, C_{\Theta \Theta}(q,z) \nonumber \\ & & + 2 \xi(z) \cos
\theta \Big\{ \left[ \left. \frac{\partial s_{00}}{\partial n} \!
\right)_T + z a_{HH}(z) \right] \cos \theta \nonumber \\ & & - z
a_{VH}(z) \Big\} C_{n \Theta}(q,z) \; .
\end{eqnarray} 
One notices that even for a general molecular fluid, transverse
current fluctuations do not couple into the $HH$ spectrum.

\subsection{Generalized Green--Kubo relations}

Ten frequency-dependent matrix elements built with the reduced
resolvent $R'(z)$ have been identified in the expressions for the
light-scattering spectra in supercooled liquids. Five generalized
transport coefficients and thermodynamic derivatives are needed in
order to describe the correlators of the hydrodynamic variables
\cite{Goetze89b}. They are the shear viscosity $K_s(z)$, the thermal
conductivity $\lambda(z)$, the dynamic specific heat $c_V(z)$, the
dynamic tension coefficient $\beta(z)$, and the longitudinal stress
relaxation kernel $K_l(z)$ (explicit expression are summarized in
Appendix C).  The remaining five frequency dependent kernels encode
the details of the light-scattering process: ${\cal S}(z)$, ${\cal
T}(z)$, $a_{VH}(z)$, $a_{VV}(z)$, $\xi(z)$.

These expressions involving reduced resolvents are very suitable for
approximations, since they do not exhibit hydrodynamic
singularities. To determine them from other theoretical approaches or
from computer simulations, however, it is more convenient to find a
formulation in terms of correlation functions involving the full
dynamics. For the considered cases at $q=0$, this is made possible by
the conservation laws which allow one to derive Green--Kubo relations
for the memory functions (transport coefficients) expressing them in
terms of (time integrals of) autocorrelation functions of the
corresponding fluxes \cite{Forster75,hansen}.  From the identity
\cite{Goetze89b}:
\begin{eqnarray} 
\label{green}
& &
(\tilde{X} | R'(z) | \tilde{Y}) = (\tilde{X}| R(z) | \tilde{Y}) +
\nonumber \\ & & (\tilde{X}|(1- R'(z) {\cal L} ) P R(z) P ( {\cal L}
R'(z) - 1) | \tilde{Y})\; ,
\end{eqnarray}
one observes that the reduced matrix elements can be rewritten as full
matrix elements and correlation functions of the hydrodynamic
variables contained in $P R(z) P $ with frequency dependent
coefficients.
Since for $q \to 0$ the coefficients involving $L n_q$ and $L j_q$
vanish due to particle and momentum conservation, only the temperature
fluctuations contribute to the frequency dependence of the
coefficients. To derive generalized Green Kubo relation, we only need
variables $\tilde{X}=QX$ and $\tilde{Y}=QY$, respectively. Therefore
all static couplings $\tilde{X}|P R(z) P $ to the hydrodynamic
variables in (\ref{green}) will vanish too and
the generalized Green-Kubo relation is given by
\begin{eqnarray}
\label{GK}
& & (X| R'(z) | Y) = (Q X| R(z)| Q Y) + \nonumber \\ & & \frac{[(X | Q
e^P) + z (X | R'(z) | e^P) ]
[ (Q e^P | Y) + z ( e^P | R'(z) | Y)] }{N T z c_V(z)} \nonumber\\ & &
\fur q\to0\, .
\end{eqnarray}
Here we made use of Eq. (\ref{energy}) and (\ref{52}) in the limit $q
\to 0$ and of the identity $R'(z) L Q = Q + zR'(z)$.  Rotational
invariance implies that the second term in \gl{GK} is nonzero only for
scalar variables $X, Y$.  Thus, e.g. the standard Green--Kubo
relations for the shear viscosity, $X=Y=\tau_{20}$, and heat
diffusion, $X=Y=\vek{j}_e$, are found. Moreover, it follows that in
the elasto-optic constant $a_{VH}(z)$, \gl{e10}, and in the tensor
background spectrum $ {\cal T}(z)$, \gl{e5}, the reduced resolvent can
be replaced with the full dynamics.

In order to obtain tractable expressions for the remaining kernels and
make contact with the standard Kadanoff-Martin approach
\cite{Kadanoff}, we define another, the conventional, fluctuating
temperature by $c_V \tilde{T}(\vek{q}) = e(\vek{q}) - n(\vek{q}) (e|
n)/(n|n)$, with normalization $( \tilde{T}(\vek{q}) |
\tilde{T}(\vek{q}) ) = NT/c_V$.  Also, let $\tilde{Q}$ denote the
projector orthogonal to density, currents and
$\tilde{T}(\vek{q})$. Choosing $X= Y = e^P$ in \gl{GK} one finds
\begin{equation}
\label{cV0}
\frac{(Q e^P(z)| Q e^P) }{NT} = \frac{c_V(z)-c_V}{z} -
\frac{(c_V(z)-c_V^0)^2}{z c_V(z)} \; .
\end{equation}
Since $Q e^P = \tilde{Q} e^P + (c_V - c_V^0) \tilde{T}$ the left-hand
side implicitly contains hydrodynamic poles due to energy
conservation. However,
\begin{equation}
\label{cV1}
\frac{(Q e^P(z) | Q e^P) }{NT} = \frac{(\tilde{Q} e^P(z) | \tilde{Q}
e^P) }{NT} - \frac{(c_V-c_V^0)^2}{z c_V} \; ,
\end{equation}
and, according to Kadanoff-Martin, the first term on the right hand
side is free of poles in the limit $z\to 0$ \cite{Kadanoff}.
Combining \gls{cV0}{cV1} one derives
\begin{equation}
\label{cV2}
c_V(z) = c_V \frac{(c_V^0)^2}{(c_V^0)^2 - z c_V (\tilde{Q} e^P(z)|
\tilde{Q} e^P)/(N T)} \; .
\end{equation}
In the liquid phase the dynamic specific heat attains its
thermodynamic value $c_V(z) \to c_V$ for $z\to 0$. Equation
(\ref{cV2}) demonstrates explicitly that the G{\"o}tze-Latz resolvent
$R'(z)$ indeed devoids all hydrodynamic singularities and is
compatible with the conventional Kadanoff-Martin formalism. It differs
in an explicit frequency dependence of $c_V(z)$, which arises from the
splitting of the conventional temperature fluctuations,
$\tilde{T}(q,t)$ into fast, the kinetic ones $\Theta(q,t)$, and
structural slow ones.

Similarly, substituting $X= p, Y= e^P$ in \gl{GK} yields
\begin{eqnarray}
& & \frac{(Q p(z) | Q e^P) }{N m T} = \frac{\beta(z)-\beta}{z} -
\frac{[\beta(z)-\beta^0] [c_V(z)-c_V^0]} {z c_V(z)} \, ,
\end{eqnarray}
where $ \beta^0 = (p | Q_n e^K)/(N m T)$. Since $Q p = \tilde{Q} p + m
\beta^0 \tilde{Q} e^P/c_V^0 + m (\beta -\beta^0) \tilde{T}$ the left
hand side can be written as
\begin{eqnarray}
& & \frac{1}{N m T} (Q p(z) |Q e^P) = \frac{1}{N m T} (\tilde{Q} p(z)
| \tilde{Q} e^P) \nonumber \\ & & + \frac{\beta^0}{N T c_V^0}
(\tilde{Q} e^P(z) | \tilde{Q} e^P) - \frac{ (\beta -\beta^0 )
(c_V-c_V^0)}{z c_V} \, .
\end{eqnarray}
Again the memory kernels appearing on the right-hand side are regular
in the low-frequency limit. Collecting terms leads to the generalized
Green-Kubo formula for the dynamic tension coefficient
\begin{equation}\label{beta2}
\beta(z) = \beta \frac{c_V(z)}{c_V} + \frac{c_V(z)}{c_V^0} z
\frac{(\tilde{Q} p(z) | \tilde{Q}e^P)}{N m T} \, .
\end{equation}

In a similar fashion the corresponding Green-Kubo formulae for the
dynamic temperature coupling, the scalar background spectrum, the
remaining term contributing to $a_{VV}(z)$, and the longitudinal
stress-stress correlation function can be obtained:
\begin{equation}\label{xi2}
\xi(z) = \left. \frac{\partial s_{00} }{\partial T} \!\! \right)_n
\frac{c_V(z)}{ c_V} + \frac{c_V(z)}{c_V^0} z \frac{(\tilde{Q}
s_{00}(z) | \tilde{Q}e^P)}{N T} \, ,
\end{equation}

\begin{eqnarray}
& & {\cal S}(z) = \frac{N T \xi(z)^2 }{z c_V(z)} - \frac{N T}{z c_V}
\left. \frac{\partial s_{00}}{\partial T} \!\! \right)_n^2
 + (\tilde{Q} s_{00}(z) | \tilde{Q} s_{00}) \, ,
\end{eqnarray}

\begin{eqnarray}
 \frac{(p | R'(z) | s_{00})}{N}& = & \frac{ m \beta(z) T}{z c_V(z)}
\xi(z) - \frac{m \beta T }{z c_V} \left. \frac{\partial
s_{00}}{\partial T} \!\! \right)_n \nonumber \\ & & + \frac{(\tilde{Q}
p(z) | \tilde{Q} s_{00})}{N} \, ,
\end{eqnarray}

\begin{equation}
\label{Kl}
\frac{K_l(z)}{n m} = \frac{m T}{z c_V(z)} \beta(z)^2 - \frac{ m T}{z
c_V} \beta^2 + \frac{(\tilde{Q} \tau^L(z) | \tilde{Q} \tau^L) }{Nm }
\, .
\end{equation}
In particular, the last term on the right-hand side of \gl{Kl} is
related to the the longitudinal viscosity $ i \eta_l = (\tilde{Q}
\tau^L(z \to 0) |\tilde{Q} \tau^L) n/N$. Therefore in the
low-frequency limit where $c_V(z) = c_V + i z c_{V}'', \beta(z) =
\beta + i z \beta''$ one finds
\begin{equation}
\label{Martin}
\frac{K_l(z\to 0)}{n m} = - i m T \beta^2 \frac{c_V''}{c_V^2} + 2 i m
T \beta \frac{\beta''}{c_V} + i \frac{\eta_l}{n m} \, .
\end{equation} 

To summarize, all ten frequency-dependent kernels can be expressed in
terms of the full resolvent $R(z)$, and therefore they can directly be
obtained from molecular-dynamics simulations.

\section{Discussion}
\label{section4}

\subsection{Depolarized spectra}

The depolarized frequency dependent spectrum, \gl{e12}, shall be
 discussed in detail as it provides the most compact expression but
 also exhibits clear qualitative changes when supercooling the liquid.
 It consists of three frequency dependent contributions, a background,
 the Pockels' constant and the transverse current correlator.

The current correlators can be taken from theories for the dynamics of
the liquid under study or from computer simulations. Alternatively,
the generalized hydrodynamic approach shifts the problem to calculate
the transverse correlator to the problem to calculate correlations of
the transversal stress tensor, viz. the frequency
dependent shear modulus,
$K_s(z)=\kuboz{\tau_{20}}{R'(z)}{\tau_{20}} n/N$, where
the resolvent devoid of hydrodynamic fluctuations from \gl{e2} appears
again. The separation of the hydrodynamic poles from structural
relaxation thus is achieved leading to \cite{Goetze89b}: \beq{e12b}
C_{jj}^T(q,z) = \frac{-(N/m)}{z+q^2 K_s(z)/nm}\; .  \eeq

The result from hydrodynamic theory for the depolarized light
scattering from equilibrium molecular liquids can be obtained if the
frequency dependence of memory functions built with the reduced
resolvent $R'(z)$ is neglected by using the Markovian low frequency
limit. Then the depolarized Pockels' constant becomes purely
imaginary: \beq{e11}
a_{VH}(z) \to i a''_{VH} = i \int_0^\infty dt \; a_{VH}(t) \fur z \to
0\; .  \eeq Therefore, and using the standard hydrodynamic result for
the shear viscosity, $K_s(z\to0)\to i \eta_s$, one recovers the result
for the depolarized spectrum first obtained within simple models in
Refs. \cite{Andersen71,Keyes71}:
\begin{eqnarray}\label{e13}
& & I_{VH}(q,\theta,\omega)_{\rm hy. l.} =  {\cal
  T}''(\omega=0) \nonumber \\ & & - q^2 \, (a''_{VH})^2 \, \cos^2
  \frac{\theta}{2} \; \frac{N q^2\eta_s/(n m^2)}{\omega^2+(q^2\eta_s/
  n m)^2}\; .
\end{eqnarray}
The transverse momentum diffusion cuts a central line of half--width
$q^2\eta_s/(mn)$ and amplitude proportional to
$(a''_{VH}\cos\frac{\theta}{2})^2nN/\eta_s$ out of a flat background.
Note, that within hydrodynamics the background is structureless, and
that the spectrum is positive owing to an elementary Schwartz
inequality, $(t_{20}|t_{20})(\tau_{20}|\tau_{20})\ge
(\tau_{20}|t_{20})^2$.

The structural relaxation of liquids cooled down to and below the
melting temperature slows down strongly and the frequency dependence
of the memory functions with reduced resolvent cannot be neglected
anymore; for a discussion of structural relaxation see e. g. the
review \cite{gs}. The result in \gl{e12} can handle this situation, as
the memory functions built with $R'(z)$ --- there are 3 in
\gls{e12}{e12b} --- may either be modeled appropriately or can be
taken from other theories or simulations.  Within the generalized
hydrodynamic approach a glass, or amorphous solid, is obtained
whenever a structural relaxation process, with time scale
$\tau_\alpha$, is slow compared to the hydrodynamic
frequencies. Assuming further that the dynamics in $R'(t)$ at shorter
times, denoted by $\tau_\beta$, can be neglected for this frequency
range, then the memory functions in \gl{e12} can be approximated by:
\beq{e14} K_s(z) \to \frac{- G_\infty}{z} + i \Gamma_s \;,\quad
a_{VH}(z) \to \frac{-a_{VH}^\infty}{z} \; , \eeq for $1/\tau_\alpha
\ll |z| \ll 1/\tau_\beta$. Note, that this is equivalent to
time--independent values, $K_s(t) = G_\infty$ and $a_{VH}(t) =
a_{VH}^\infty$ for $\tau_\beta\ll t\ \ll\tau_\alpha$. Therefore, the
poles in \gl{e14} are called non--ergodicity poles as they describe
frozen--in, non--relaxing components. $G_\infty = m n c_T^2$ is the
glassy shear modulus familiar from Maxwell's model and $a_{HV}^\infty$
is the Pockels' constant (often denoted ${\cal P}_{44}$) quantifying
the elasto--optic coupling in the glass \cite{Landau8}. Whereas
$G_\infty$ and $\Gamma_s$ need to be positive, the sign of
$a_{HV}^\infty$ is undetermined; an imaginary part in $a_{HV}$ exists 
in principle but does not contibute to the spectrum in the hydrodynamic limit.
 Equations (\ref{e12},\ref{e14}) predict for the hydrodynamic glass spectrum:
\begin{eqnarray}\label{e15} 
& & I_{VH}(q,\theta,\omega)_{\rm hy. g} =  {\cal T}_g'' +
\nonumber\\ & & (q \cos \frac{\theta}{2} \, a_{VH}^\infty)^2 \;
\frac{q^2\Gamma_s/(m n)}{(\omega^2-q^2c_T^2)^2+ (\omega q^2
\Gamma_s/nm)^2}\; .
\end{eqnarray}
Two transverse phonon peaks characterized by the transverse sound
velocity, $c_T$, and a width $\propto q^2 \Gamma_s$, appear which are
described as damped harmonic oscillations. The background consists of
a central line which cannot be resolved and a structureless continuum,
${\cal T}(z) = -{\cal T}_\infty/z+ i {\cal T}_g''$.

Note, that both hydrodynamic expressions, \gls{e13}{e15}, violate the
sum--rule for the total intensity, \gl{e9c}, and predict wave vector
dependent total scattered intensities. The reason, of course, are the
Markovian approximations in \gls{e12}{e14} which fail to describe the
dynamics outside the hydrodynamic range. Non--trivial spectra obtained
in glasses on frequency scales characterized by $\tau_\beta$
\cite{Surovtsev98} also require more elaborate expressions for the
memory functions in \gl{e12}.

For temperatures around the liquid--to--glass crossover at $T_c$,
neither the assumption $\tau_\alpha \omega \gg 1$, nor the estimate
$\tau_\beta \omega \ll 1$ hold and the depolarized spectra exhibit
anomalies \cite{Tao91}.  The mode coupling theory of the structural
relaxation there suggests to model the reduced resolvent as $K_s(z)
\approx -G_\infty [ 1 - (1-iz \tau_\alpha)^{-\beta_{CD}} + (-i z t_0)^a
]/z$ for $T\approx T_c$, and similar expressions for the other two
memory functions. Whereas the Cole--Davidson behavior (the first two
terms) is a (rough) model of the $\alpha$--process, and $\beta_{CD}$ will
differ for different resolvent matrix elements, the 
power law with exponent $a$ describes the universal ``critical'' decay
close to the transition. Here $t_0$ is a microscopic time and the
exponent $a$ and the (true, universal) exponent $b$ of the high
frequency von Schweidler wing of the $\alpha$--process,
$K_s(1/\tau_\alpha\ll z \ll 1/\tau_\beta) \sim (-\ii
z\tau_\alpha)^{-b}/z$, are related; see e. g. the review \cite{gs} for
further information.

\subsection{Polarized spectra}

A most prominent feature of the polarized spectrum is the Brillouin
peak arising from propagating sound waves. Upon cooling the liquid,
structural relaxation manifests itself predominately by a gradual
change of the sound velocity and the damping constant. Considering the
enormous increase of the transport coefficients, e.~g. the
longitudinal viscosity which describes the damping in the liquid, this
clearly points out the necessity to consider the frequency dependence
of the reduced resolvent as done in generalized hydrodynamcis.
Furthermore, the couplings $a_{VV}(z), \xi(z)$ as well as the
background spectra ${\cal S}(z), {\cal T}(z)$ exhibit non-trivial
$z$-dependences in the frequency regime of interest.

The structural relaxation of the Pockels' constant $a_{VV}(z)$ (with
 numerically different constants) can be modelled as given in Eqs.
 (\ref{e11}) and (\ref{e14}) and as described at the end of the
 previous section.  The explicit factor $z$ in the frequency dependent
 part in $\xi(z)$, \gl{xi}, cancels a possible non--ergodicity
 pole. The coupling to temperature fluctuations interpolates smoothly
 between its low-frequency thermodynamic value $(\partial
 s_{00}/\partial T)_n$ and a high-frequency coupling, $\xi_\infty$,
 characteristic
 for a glass.  A renormalization also appears in the effective
 coupling to the density fluctuations, which in \gl{VVspectrum} is
 described by the Pockels' constant ${\cal P}_{12}=(\partial
 s_{00}/\partial n)_T + z a_{VV}(z)$ and in \gl{HHspectrum} by $({\cal
 P}_{12} \cos\theta-{\cal P}_{44})$, respectively, where ${\cal
 P}_{44}=z a_{VH}(z)$.
  Note, that the frequency dependent
 renormalization of $(\partial s_{00}/\partial T)_n$ in \gl{xi}
 vanishes, if only the hydrodynamic fluctuations, density and
 temperature, contribute to the scalar scattering; then also
 $a_{VV}(z) = \frac 23 a_{HV}(z)$.

In order to obtain the spectrum in the true hydrodynamic limit one
substitutes the appropriate correlation functions ---   see Appendix C ---
and replaces all  memory functions with their low frequency  limits; only 
in $c_V(z)$, $\beta(z)$ and $\xi(z)$ linear terms in $z$ need to be kept as 
can be seen from Eqs. (\ref{cV2},\ref{beta2},\ref{xi2}).
There are the three familiar hydrodynamic resonances
superimposed on the Raman background: the Brillouin doublet of sound
modes and the Rayleigh heat pole. 
The spectrum is obtained from determining the residues of these poles to
lowest order in frequency and wave vector. 
\begin{eqnarray} \label{e42}
& &
I_{VV}(q,\theta,\omega)_{hy.l.}  = 
(\tilde{Q} s_{00}(\omega=0) | \tilde{Q}  s_{00})''
+ 4 {\cal T}''(\omega=0)/3 + \nonumber \\ & & N n \kappa_T 
{\cal P}_{12}^2 
 \Big[\frac{X^{\rm Br.}}{\gamma} \frac{c^2 q^4
\Gamma_l}{(\omega^2-c^2 q^2)^2 + (\omega q^2 \Gamma_l)^2} \nonumber \\
& & + X^{\rm R.} 
\frac{\gamma-1}{\gamma} \frac{q^2 D_T }{\omega^2 + (q^2 D_T)^2}
\Big] \; ,
\end{eqnarray}
where the adiabatic sound velocity $c=\sqrt{\gamma/(mn\kappa_T)}$, the
 longitudinal sound damping $\Gamma_l=D_T(\gamma-1)+\eta_l/(mn)$, and
 the heat diffusion constant $D_T=\lambda''/ c_P$ appear (see Appendix
 C for details). The Pockels constant is given by the thermodynamic derivative,
${\cal P}_{12}=\partial s_{00}/\partial n)_T$, and
 the flat background consists of scalar and tensor  parts.
Neglecting contributions from temperature fluctuations, 
$X^{\rm R.}= X^{\rm Br.}=1$,  one regains the
well--known result for light scattering from (hydrodynamic)
density fluctuations \cite{Landau8,Forster75,Berne76}. Then, the
Landau--Placzek result $(\gamma-1)$ is recovered for the relative
intensity of the Rayleigh to the Brillouin lines, where
$\gamma=c_P/c_V$ denotes the ratio of the isobaric to isochoric heat
capacity \cite{Landau8,Forster75,Berne76}. In the general case, scattering
from temperature fluctuations leads to (presumably small) corrections:
$X^{\rm R.}=(1-
\frac{c_0^2}{\beta} \frac{\xi}{{\cal P}_{12}})^2$ and
$X^{\rm Br.}=(1+(\gamma-1)
\frac{c_0^2}{\beta}  \frac{\xi}{{\cal P}_{12}} )^2$, where
$\xi=\partial s_{00}/\partial T)_n$.

For the glass, $\omega\tau_\alpha\gg1$, actually the identical formula
 holds where, however, the isothermal compressibility, the 
sound velocity and damping constants, as
 well as the couplings  and the  background --- compare \gls{e13}{e15} ---
are renormalized. The high frequency values of the memory functions
appear in the formally identical definitions of $c$, $D_T$ and $\Gamma_l$,
where, in the frequency window 
$1/\tau_\alpha\ll \omega \ll 1/\tau_\beta$, simple Markovian expressions
like $c_V(z) = c_V^\infty + i z c_{V,g}''$ are appropriate for
ergodic matrix elements  ($c_V(z)$, $\beta(z)$, $\xi(z)$ 
and $\lambda(z)$), and frozen--in components, leading to 
$K_l(z) = -K_l^\infty/z+  i K_{l,g}''$, appear in $K_l(z)$, $a_{VV}(z)$ and in
the Raman background lines.
Thus, e.~g. the expression for the Pockels constant becomes:
${\cal P}_{12}^\infty=\partial s_{00}/\partial n)_T-a_{VV}^\infty$.
A non--trivial renormalization appears in the isothermal sound velocity 
or equivalently the isothermal compressibility,
because of the frozen structural relaxation in the longitudinal
friction function:
\beq{e43}
(c_T^\infty)^2 = \frac{1}{m n \kappa_T^\infty} = c_0^2 + \frac{K_l^\infty}{m n}
\; ,
\eeq
as first observed by Mountain \cite{Mountain66} and predicted from
microscopic expressions by the mode coupling theory \cite{gs}.
Note that the reduction of the Brillouin and Rayleigh intensities described by
\gls{e42}{e43} is caused by the appearance of frozen--in density fluctuations
(an elastic Mountain--line) which contributes the missing weight,
$N n \kappa_T ((\partial s_{00}/\partial n)_T^2 - (\frac{c_0}{c_T^\infty})^2 
({\cal P}_{12}^\infty)^2)$.

For the $HH$ spectrum let us just mention that for $\theta = \pi/2$,
\gl{HHspectrum} simplifes considerably, since the scalar scattering
drops out completely. For frequencies $\omega \tau_\alpha \ll 1$,
$a_{VH}(z) \to i a_{VH}''$ holds again, and one finds
\begin{equation}
I_{HH}(q,\theta= \frac{\pi}{2},\omega)_{hy.l.} = {\cal T}''(\omega =
0) - ( \omega a_{VH}'')^2 C_{nn}(q,\omega)'' \; .
\end{equation}  
The hydrodynamic modes are suppressed by a factor of $\omega^2$,
e.g. the Brillouin line cuts out a Lorentzian with half-width $q^2
\Gamma_l$ and amplitude $N (a_{VH}'')^2/(m \Gamma_l)$ out of a flat
background.

\subsection{Intensity ratios}
Interesting intensity ratios can be constructed if the scattered
intensities with different polarizations of the light before and after
the scattering process are considered.  The standard depolarization
ratio compares the off-resonance ($\omega\gg cq$) frequency-dependent
background intensities neglecting all hydrodynamic modes:
\begin{equation}\label{e19a}
 \frac{ I_{VH}(q,\theta,\omega)}{I_{VV}(q,\theta,\omega)} = \frac{
 {\cal T}''(\omega) }{ (\tilde{Q} s_{00}(\omega) | \tilde{Q}
 s_{00})'' + 4{\cal T}''(\omega)/3} \leq 3/4\; .
\end{equation}
The expected depolarization ratio $3/4$ is recovered if the only
scalar scattering mechanisms are density and temperature fluctuations.

\section{Comparison with phenomenological approaches}
\label{section5}

It appears worthwile to discuss earlier phenomenological approaches
within our general framework. Again we shall concentrate on the
depolarized spectrum which has been the focus of a long list of
theoretical descriptions.

The original approach of Andersen and Pecora \cite{Andersen71} and
Keyes and Kivelson \cite{Keyes71} captures the spectra as predicted by
hydrodynamic theory. Earlier approaches had failed to do so, because
they overly constraint the hydrodynamic transport coefficients. The
two or three variable approaches in \cite{Andersen71,Keyes71} go
beyond hydrodynamics as they model the background spectrum as a (sum
of) Lorentzian(s).  But they miss the frequency dependence of the
shear modulus and of the Pockels constants and thus do not present a
valid extension beyond the hydrodynamic result in
\gl{e13}. Consequently their results could not be applied to
supercooled liquids.  They also failed to recover the sum rule in
\gl{e9c}.  Vaucamps et al. extended this approach by using Maxwell's
model, i. e. the approximation $K_s(z)= -mnc_T^2/[z+i/\tau_s]$ for the
shear modulus \cite{Vaucamps}. More elaborate models, which identify
slower and faster relaxational processes in the shear modulus
$K_s(t)$, were suggested in
Refs. \cite{Quentrec,Wang1,Chapell}. Models with more than two
variables can be brought into a form suggested by a simple
viscoelastic approximation (VA) to our general result, \gl{e12}:
\begin{eqnarray}\label{dis1} 
I_{VH}^{\rm VA}(z) \propto \frac{-1}{z+i \Gamma_{\cal T}} +
\frac{1}{[z+i \gamma_{VH}]^2} \frac{-A}{z- q^2 c_T^2/[z+i/\tau_s]} \;
,
\end{eqnarray}
where $A= (qa_{VH}^\infty\cos\frac \theta2)^2N/(m{\cal T}^\infty)$ and
\gl{dis1} gives a simplified $\alpha$--process only description akin
to Maxwell's model.  The multi--variable phenomenological models,
however, rest on specific assumptions about the
rotational--translational coupling, and are, therfore, forced to
introduce constraints on the coupling parameters in order to recover
\gl{dis1}; see Ref. \cite{Drey2} for a detailed discussion of this
aspect.  Note that the $\alpha$--process only model, \gl{dis1}, fails
to describe higher frequencies as can be seen from its incorrect
prediction of a vanishing damping of the shear waves in the glass, see
\gl{e15}, and from its failure to recover the exact sum rule,
\gl{e9c}.

The introduction of retardation effects via memory functions in the
work of Wang \cite{Wang1} and Dreyfus et al. \cite{Drey1,Drey2}
removed constraints on the coupling parameters, and, within the latter
phenomenological approach, an expression formally equivalent to
\gl{e12} was given.

\section{Conclusion}
\label{section6}

In this paper we discussed the light scattering spectra for a
one-component molecular liquid incorporating slow structural
relaxation and thus extending the description to supercooled liquids
and glasses. In contrast to earlier phenomenological approaches no
assumptions are made on the nature of the scattering mechanisms and on
the origin of the structural relaxation. 
Our results are akin to the
theory of neutron scattering experiments where it is shown that, in
general, density fluctuation functions are measured without the need
for special models of their dynamics \cite{hansen,Forster75}. For
example, \gls{e12}{e12b} indicate how the shear viscosity can be
measured by light scattering, without specifying what microscopic
dynamical mechanisms contribute to the decay of transversal currents.
 In particular, we clarify
the question of how many generalized frequency dependent
elasto-optical constants and memory functions are needed for a general
description of the long-wave-length fluctuations of the dielectric
tensor. Furthermore we give explicit microscopic formulae for the
background spectrum and the elasto-optical constants which could serve
as a starting point for approximations once a choice for the
scattering mechanism is made.  For isotropic particles considering
dipole--induced--dipole scattering this has been performed in
\cite{Fuchs91}. Alternatively, computer simulations could be employed.
Our central assumption is to consider systems characterized by
short--ranged equilibrium correlations. Thus we can neglect the wave
vector dependence in matrix elements describing the structural
relaxation leading to Green--Kubo like formulae for the memory
functions as familiar from simple hydrodynamics.

Whereas previous approaches were mainly concerned with the nature of
the depolarized spectrum and correspondingly e. g.  attribute the
scalar fluctuations to a combination of density and temperature
fluctuations, we carefully make the distinction of dielectric
fluctuations coupled to hydrodynamic modes and the ones orthogonal
thereto.  Consequently one obtains a theory that combines conventional
Brillouin and Raman scattering. For example, molecular vibrations or
rotational motion give rise to scalar as well as tensor scattering
which appears as background to the hydrodynamic resonances and is
included in our framework.

An important aspect of our results is that no assumptions on e. g.
translational--rotational coupling or about the concrete description
of the structural relaxation were necessary. The general aspects of
the light scattering spectra were worked out and correlation functions
were defined, which can, in principle at least, be measured
experimentally, and which contain the general information about the
dynamics of the sample under study.

\acknowledgments{ Valuable discussions with Prof. W. G{\"o}tze and
Prof. H.~Z. Cummins and their helpful comments on the manuscript are
gratefully acknowledged.  This work was supported by the Deutsche
Forschungsgemeinschaft under Grants No. Fr 417/2 and Fu 309/3.}

\appendix
\section{Scattering geometries}
The scattering plane, i.e., the plane that contains both the wave
vectors $\vek{k}_i,\vek{k}_f$ of the incident and scattered wave, is
chosen as the $x$--$z$ plane. The direction of the momentum transfer
$\vek{q} = \vek{k}_i - \vek{k}_f$ is taken to be $(0,0,-q)$.  Since we
consider only small frequency shifts $|\vek{k}_i | \simeq |\vek{k}_f|
$ the scattering angle $\theta = \angle(\vek{k}_f, \vek{k}_i)$ is
related to the momentum transfer via $ q = 2 k_i \sin
\frac{\theta}{2}$. The dielectric fluctuations corresponding to the
conventional scattering geometries, viz. polarization perpendicular
($V$) to and in the scattering plane ($H$), are then given by
\cite{Berne76}:
\begin{eqnarray}
\delta \epsilon_{VV}(\vek{q}) & = & \delta \epsilon_{yy}(\vek{q})
\nonumber \\ \delta \epsilon_{VH}(\vek{q}) & = & \delta
\epsilon_{xy}(\vek{q}) \sin \frac{\theta}{2} - \delta
\epsilon_{yz}(\vek{q}) \cos \frac{\theta}{2} \nonumber \\ \delta
\epsilon_{HV}(\vek{q}) & = & \delta \epsilon_{xy}(\vek{q}) \sin
\frac{\theta}{2} + \delta \epsilon_{yz}(\vek{q}) \cos \frac{\theta}{2}
\nonumber \\ \delta \epsilon_{HH}(\vek{q}) & = & \delta
\epsilon_{xx}(\vek{q}) \sin^2 \frac{\theta}{2} - \delta
\epsilon_{zz}(\vek{q}) \cos^2 \frac{\theta}{2}
\end{eqnarray}
The scattering intensities are expressed via the spectra of the
dielectric fluctuations
\begin{eqnarray}
I_{io}(q,\theta,\omega) & = & \left( \epsilon_{io}(\vek{q},\omega) |
\epsilon_{io}(\vek{q}) \right)'' \nonumber \\ & = & \int_0^\infty dt
\cos (\omega t) \left( \epsilon_{io}(\vek{q},t) |
\epsilon_{io}(\vek{q}) \right)
\end{eqnarray} 
The scattering intensities obey the sum rule
\begin{equation}
\frac{1}{2\pi} \int_0^\infty d \omega I_{io}(q,\theta,\omega) =
I_{io}(q,\theta)
\end{equation}
where the total intensities $I_{io}(q,\theta)$ are determined by the
thermal fluctuation of the dielectric tensors
\begin{equation}
\label{total}
I_{io}(q,\theta) = \left( \epsilon_{io}(\vek{q},t=0)|
\epsilon_{io}(\vek{q}) \right) = \frac{\langle |\delta
\epsilon_{io}(\vek{q})|^2 \rangle }{k_B T} \; .
\end{equation}

\section{Polar fluids}

Due to the long--ranged character of dipolar interactions, fluids
of molecules possessing a permanent (electric or magnetic) dipole
moment need additional considerations. Surface polarizations (in the
case of electric dipole moments) or demagnetizing fields (for magnetic
moments, where we will use the electric language in the following)
arise for an {\it uniform} sample and lead to a dielectric response to
the applied external (vacuum) field which depends on the shape of the
sample \cite{hansen,Banerjee98}.  Thus our use of rotational isotropy
in the main text appears not justified.  However, because of the
strong angular dependence of the dipolar interaction, polar fluids can
lower their free energy by the formation of domains or textures, and
macroscopically isotropic response functions and dielectric
fluctuations can be expected \cite{Banerjee98,Deutch73,Wertheim79}.
For these situations, where ferroelectric order is absent, and where
``chain--formation'' \cite{Rosensweig85} does not lead to large
aggregates, i.~e. where $qa\ll1$ holds, our general formulae apply.

\section{Generalized hydrodynamics}
\label{genHyd}
The correlation functions in the generalized hydrodynamic limit can be
expressed as \cite{Goetze89b}
 \begin{eqnarray}
C_{nn}(q,z) & = & - N n \kappa_T /[ z- (c_0 q)^2 / [ z+ q^2 K_l(z)/(m
n) \nonumber \\ & & - q^2 m T \beta(z)^2 / [ z c_V(z) + q^2
\lambda(z)]]] \; ,
\end{eqnarray}
\begin{eqnarray}
& & C_{n\Theta}(q.z) = -N T q^2 \beta(z) /[ [z^2-(c_0 q)^2 \nonumber
\\ & & + z q^2 K_l(z)/(m n)] [ z c_V(z) +q ^2 \lambda(z)] - q^2 m T
z\beta(z)^2 ] \; ,
\end{eqnarray}
\begin{eqnarray} \label{52}
& & C_{\Theta \Theta}(q,z) = - NT /[ z c_V(z)+ q^2 \lambda(z)
\nonumber \\ & & - q^2 m T z \beta(z)^2 /[ z^2 -(c_0 q)^2 +z q^2
K_l(z)/(m n)]] \; .
\end{eqnarray}
Here $c_0 = ( m n \kappa_T)^{-1/2}$ denotes the isothermal sound
velocity, $K_l(z) = (\tau^L | R'(z) | \tau^L) n/N$ the longitudinal
stress relaxation kernel, and $\lambda(z) = (j^L_e | R'(z) | j^L_e )/
(N T)$ the thermal conductivity. The frequency-dependent tension
coefficient and specific heat are given by
\begin{eqnarray}
\label{beta}
\beta(z) & = & \beta + z (p | R'(z) | e^P )/(N m T) \; ,
\end{eqnarray}
\begin{eqnarray}
\label{cV} 
c_V(z) & = & c_V + z (e^P | R'(z) | e^P)/(N T) \; ,
\end{eqnarray}
where $\beta = (p | Q_n e)/(N m T), \, c_V = (e | Q_n e)/( N T)$
retain their standard values.

There appear three hydrodynamic modes in the liquid state.  First, the
sound doublet $z = \pm c q - \ii q^2 \Gamma_l/2 +{\cal O}(q^3)$ with
the adiabatic sound velocity, $c^2 = c_0^2 + m T \beta^2/c_V$ and
\begin{equation}
\Gamma_l = \frac{K_l''}{n m} + \frac{m T \beta^2}{c_V} [
\frac{\lambda''}{c^2 c_V} + \frac{c_V''}{c_V} - 2
\frac{\beta''}{\beta}] \; ,
\end{equation}
where we have written $\ii K_l'' = K_l(z\to 0), \ii \lambda'' =
\lambda(z\to 0), c_V(z\to 0) = c_V + \ii z c_V'', \beta(z\to 0) =
\beta + \ii z \beta''$.  In the generalized hydrodynamic approach the
sound damping consists of four parts: viscous friction, losses at
conversion from mechanical into thermal energy, losses from storing
and extracting thermal energy, and thermal diffusion.  Although the
expression looks unfamiliar, the associated Green-Kubo formula,
\gl{Martin}, shows that the longitudinal viscosity appears, leading to
the familiar result \cite{Landau8,Forster75}.  Second, there is the
heat mode $z= -\ii q^2 D_T$, where the thermal diffusion constant
reads
\begin{equation}\label{heat}
D_T = \frac{c_0^2}{c^2} \frac{\lambda''}{c_V} \; =
\frac{\lambda''}{c_P}\; .
\end{equation}

For the second equality in (\ref{heat}) we made use of standard
thermodynamic transformation formulas.  The merit of the generalized
hydrodynamic approach lies in the separation of the long-wavelength
and low-frequency properties \cite{Goetze89b}. The only wave-vector
dependence arises from the conservation laws.  The structural
relaxation is captured in generalized transport and thermodynamic
derivatives, which exhibit a significant frequency-dependence in the
regime of interest.

\section{Special scattering mechanisms}

In order to show that all contributions to the light scattering
spectra discussed in the text will be present in general, and also in
order to exemplify how our general formulae can be used, specific
scattering mechanisms are listed:

1. If single scattering processes from biaxial (likely chiral)
   molecules are considered, the background spectrum arises from the
   two irreducible spherical tensors:
\begin{equation}\label{ala}
s_{00} = \frac{1}{3} (\alpha_{xx}+\alpha_{yy}+\alpha_{zz}) n(\vek{q})
\, , \quad q \rightarrow 0\; ,
\end{equation}
\begin{eqnarray*}
t_{2\nu} &=& \frac{1}{\sqrt{8}} (\alpha_{xx}-\alpha_{yy}) \sum_i [
D^{(2)}_{\nu,2}(\Omega_i)+D^{(2)}_{\nu,-2}(\Omega_i)] \\ & & +
\frac{1}{\sqrt{12}} (2 \alpha_{zz}-\alpha_{xx}-\alpha_{yy}) \sum_i
D^{(2)}_{\nu,0}(\Omega_i)\; ,
\end{eqnarray*}
where the scalar dynamics follows the density fluctuations and and the
Wigner functions $D^{(j)}_{\nu \mu}(\Omega)$ capture the dynamics of
the molecular orientation in terms of Euler angles $\Omega =
(\varphi,\vartheta, \chi)$ \cite{Gray}.
 
2. In the special case of linear molecules, $\alpha_{xx} =
 \alpha_{yy}$, the tensorial fluctuations simplify
\begin{eqnarray}
t_{2\nu} &=& \sqrt{\frac{4\pi}{15}} ( \alpha_{zz}-\alpha_{xx}) \sum_i
Y_{2\nu}(\theta_i,\varphi_i)\;,
\end{eqnarray}
and the standard spherical harmonics appear \cite{Berne76}.

3. In the case of optically isotropic particles, depolarized
   scattering can arise from first order dipole induced dipole (DID)
   scattering \cite{Gelbart74}:
\begin{equation}
  t_{2\nu} = \frac{\alpha^2}{\sqrt{2}} \int \frac{d^3 k}{(2\pi)^3} n(-{\bf k})
  T_{2\nu}(\hat{\bf k}) n({\bf k})
\end{equation}
where $T_{2\nu}(\hat{\bf k}) = -4 \pi\sqrt{8 \pi/15} Y_{2\nu}(\hat{\bf
k})$ is the static dipole tensor.

4. If second order DID scattering of isotropic particles is
   considered, then the scalar scattering consists of a contribution
   coupling to the density fluctuations, \gl{ala}, and of another
   non--hydrodynamic contribution:
\begin{eqnarray}
& & s_{00} = \frac{\alpha^3}{3} \sum_\mu \int \! \frac{d^3 k \; d^3
p}{(2\pi)^6} \; (-1)^\mu \times \nonumber \\ & & T_{2,\mu}(\hat{\bf
k}) T_{2,-\mu}(\hat{\bf p}) n(-{\bf k}-{\bf p}) n({\bf k}) n({\bf
p})\; ,
\end{eqnarray}
\begin{eqnarray}
& & t_{2\nu} = -\sqrt{\frac{35}{24}} \alpha^3 \sum_\mu (-1)^\nu \left(
\begin{array}{ccc} 
2 & 2 & 2 \\ \mu & \nu-\mu & -\nu \\
\end{array}
\right) \times \nonumber \\ & & \int \! \frac{d^3 k \; d^3 p}{(2
\pi)^6} \; T_{2,\nu-\mu}(\hat{\bf k}) T_{2,\mu}(\hat{\bf p}) n(-{\bf
k}-{\bf p}) n({\bf k}) n({\bf p})\; .
\end{eqnarray}
This implies that in this case the depolarization ratio from \gl{e19a}
is smaller than the expected value $3/4$. Furthermore the scalar
fluctuations overlap with the orthogonalized energy, i.e. $(s_{00} |
Q_n e)$ is nonzero, and the coupling of dielectric fluctuations to
temperature are relevant.  In particular, one finds that even in the
hydrodynamic limit for a liquid the Landau-Placzek ration is not
fulfilled.

5. The lower depolarization ratio also already arises from first order
   DID of linear molecules:
\begin{eqnarray}
s_{00} & = & \frac{2}{3} \alpha (\alpha_{zz}-\alpha_{xx}) \times
 \nonumber \\ & & \sum_\mu (-1)^\mu \int \! \frac{d^3 k}{(2 \pi)^3} \;
 T_{2,-\mu}(\hat{\bf k}) n(-{\bf k}) n_{2\mu}({\bf k})
\end{eqnarray}
\begin{eqnarray}
t_{2\nu} & = & -\sqrt{\frac{35}{6}}\alpha (\alpha_{zz}-\alpha_{xx})
\sum_\mu \int \! \frac{d^3 k}{(2 \pi)^3} \; (-1)^\nu \times \nonumber
\\ & & \left(
\begin{array}{ccc}
2 & 2 & 2 \\ \mu & \nu-\mu & -\nu \\
\end{array} \right)
T_{2,\nu-\mu}(\hat{\bf k}) n(-{\bf k}) n_{2\mu}({\bf k}) \; ,
\end{eqnarray}
where the fluctuating tensor density reads $n_{2 \mu}(\vek{k}) =
\sqrt{8 \pi/15} \sum_i Y_{2\mu}(\theta_i,\varphi_i) \exp(-\ii \vek{k}
\cdot \vek{r}_i)$.

6. If the molecules are not considered as rigid there may be Raman
 active modes due to intramolecular vibrations. If $Q_{a,i}$ is the
 vibrational coordinate and $(\partial \alpha/\partial Q_a)$ the
 corresponding change of the polarizability for mode $a$ there is a
 contribution to the scalar dielectric fluctuations
\begin{equation}
 s_{00} = \sum_{i} \sum_a \frac{\partial \alpha}{\partial Q_a} Q_{a,i}
 \,
\end{equation}
Here the sums run over all molecules and Raman modes. A similar
expression can be derived for the dielectric tensor fluctuations.

\end{multicols}

\end{document}